\documentclass[]{iopart}
\usepackage[top=1.1in, bottom=1in, left= 1.1in, right= 1in]{geometry} 	
\usepackage{graphicx}
\usepackage{bm}
\usepackage{color}

\begin{document}
\fontsize{10.8pt}{13.8pt}\selectfont
\title[Effects of Flow Collisionality on ELM Replication in Plasma Guns]{Effects of Flow Collisionality on ELM Replication in Plasma Guns}
\author{Thomas C. Underwood$^{1}$, Vivek Subramaniam$^{2}$, William M. Riedel$^{1}$, Laxminarayan L. Raja$^{2}$, and Mark A. Cappelli$^{1}$}
\address{$^{1}$ Department of Mechanical Engineering, Stanford University, CA}
\address{$^{2}$ Department of Aerospace Engineering and Engineering Mechanics, The University of Texas, Austin, TX}
\ead{tunderw5@stanford.edu}

\begin{abstract}

Degradation of first wall materials due to plasma disturbances severely limit both the lifetime and longevity of fusion reactors. Among the various kinds of disturbances, type I edge localized modes (ELMs) in particular present significant design challenges due to their expected heat loading and relative frequency in next step fusion reactors. Plasma gun devices have been used extensively to replicate ELM conditions in the laboratory, however feature higher density, lower temperatures, and thus higher flow collisionality than those expected in fusion conditions. This work presents experimental visualizations that indicate strong shocks form in gun devices over spatial and temporal scales that precede ablation dynamics. These measurements are used to validate detailed magnetohydrodynamic simulations that capture the production of plasma jets and the shielding effect collisionality plays in particle transport to material surfaces. Simulations show that self-shielding effects in plasma guns reduce the free streaming heat flux by up to 90\% and further reduce the incoming particle kinetic energy impinging on material surfaces. These simulations are performed over a range of operating conditions for gun devices and a discussion is provided regarding how existing experimental measurements can be interpreted when extrapolating to fusion conditions.

\end{abstract}

\section{Introduction}

Loss of stable confinement within magnetic fusion reactors causes significant damage to plasma facing components (PFC) and in doing so, severely limits both the feasible design space and operational lifetime of conventional tokamaks. In next step fusion devices, off-normal events including major disruptions, vertical displacement events, and edge localized modes (ELMs) all pose significant challenges to candidate materials as ablation, melting, and vaporization are expected \cite{Leonard1999}. During such events, both energy and particles are transported from the confined plasma first into the scrape-off layer (SOL) and eventually onto the first wall and divertor surfaces to relieve pressure gradients and relax the system back to equilibrium. Among these, ELMs are particularly concerning because of their relative frequency and role in providing both particle and impurity control in operating reactors \cite{Federici2001}. This problem is exasperated in next step devices, such as ITER, where gain and energy confinement criteria require operation in the type I ELMy H-mode. A characteristic feature of this operational regime is the type I ELM disruption where between 3-10\% of the stored core energy is expelled periodically into the SOL over a period spanning 0.1-1.0 ms. Of this energy, 50-80\% is expected to be directed toward the diverter targets resulting in peak heat fluxes of 1-10 GW/m$^{2}$ \cite{Loarte2007}. The exact origin of these events and where and how this energy is transferred to PFCs is an important area of active research that is critical for prolonged operational success of fusion devices.

A detailed characterization of ELMs is challenging in tokamaks where operational costs and limited diagnostic access restrict the interaction physics that can be uncovered. In addition, the magnitude of the disruption conditions expected in ITER cannot be achieved in existing tokamaks \cite{Federici2001}. To fill this gap, laboratory experiments including lasers, electron beams, and plasma guns have been used extensively to simulate disruption events. For example, both lasers and electron beams have been used to test divertor materials and changes to surface morphology \cite{Nakamura1992}. In the case of laser ablation however, the small beam size ($\leq 2$-4 mm) and large photon penetration depth are inconsistent with the deposition physics expected in ELMs. Similar issues arise in electron beams where high-energy (100-150 keV) electrons easily penetrate ablated material and eliminate a source of shielding that appears in plasma driven thermal transport \cite{Hassanein1994}. Plasma guns have been used to study both pulsed \cite{Safronov2001,Arkhipov1996,Belan1996,Chebotarev1996,Crawford1993}  and steady-state heat \cite{Temmerman2011} heat loading to candidate materials. They are thought to more realistically simulate disruption events as a high-velocity quasineutral plasma jet supplies the energy flux to materials, consistent with those found in ELMs.

Plasma gun experiments as a whole have contributed significantly to quantifying both the ablation and degradation effects of candidate first wall materials. They have been used to uncover the crack pattern and residual stresses that form in tungsten targets under repetitive thermal loading \cite{Garkusha2009v2} and further investigate erosion mechanisms in a variety of candidate materials \cite{Garkusha2009}. They have also been instrumental in studying the vapor shielding process where a thin layer of material is ablated and forms a protective layer over the surface to limit subsequent thermal transport. Although energy densities and heat fluxes similar to type I ELMs ($\geq 1$ GW/m$^{2}$) are achievable in such devices, the characteristic densities and temperatures are orders of magnitude different than those found in operating fusion reactors. This results in dramatically different ion kinetic energy and increased collisionality compared to ELM conditions expected in fusion reactors. What remains is to quantify the effect this added collisionality has on the effective energy transfer and degradation rates of candidate materials.

Accurate simulation of ELMs in the laboratory require matching both the bulk disruption features and the plasma properties that make them up. For plasma guns specifically, extensive experimental measurements of thermal loading and broader material damage mechanisms have been measured in conditions mimicking ELMs. Moreover, detailed simulations involving target thermodynamics, hydrodynamics, and surface chemistry in addition to vapor, melt layer, and sheath effects have been used to isolate underlying physics in the degradation process. This paper acts to fill in the gap between ELMs generated in fusion conditions and the plasma guns that simulate them \cite{Hassanein1994,Hassanein1999}. We build upon the prior numerical work detailing the plasma-material coupling process and focus on capturing the generation of plasma jets in gun devices and how they stagnate on material targets. Experimental evidence of a significant added shock shielding mechanism unique to plasma guns is presented and further used to validate simulations. Finally, we quantify the effect that this shock has on both the energy densities and heat flux reaching material surfaces and specifically discuss how plasma gun results can be interpreted given this additional source of shielding.

\section{Plasma Guns}

\begin{figure*}[tp]
\centering
\includegraphics[width=0.7\textwidth]{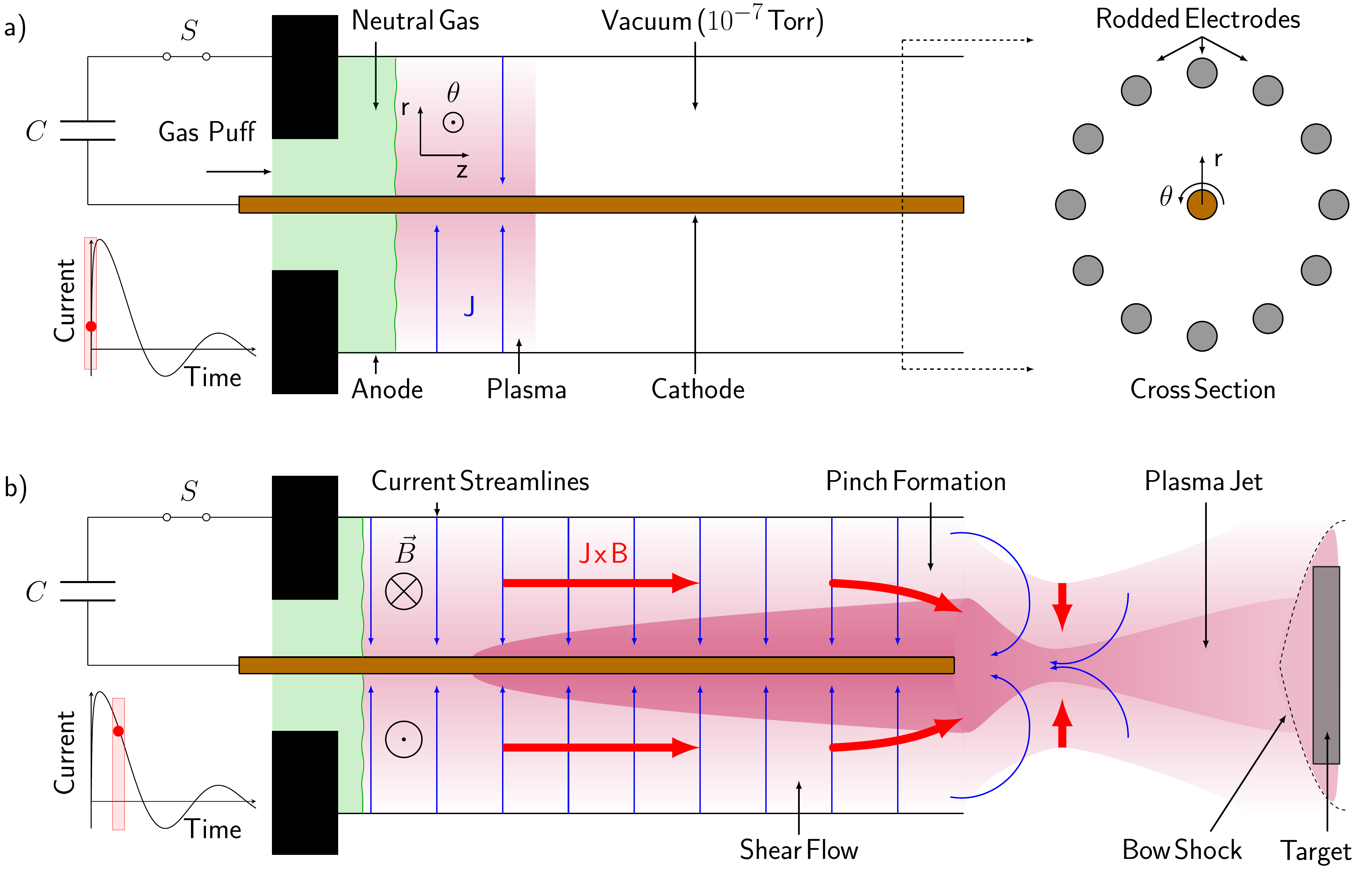}
\caption{Schematic showing the operational principles of a plasma gun. Specifically, (a) details the electrode geometry for the Stanford facility and the process by which a plasma jet is generated, and (b) shows how the plasma jet interacts with a material target.}
\label{fig:theory}
\end{figure*}

\subsection{Operational theory}

Plasma guns produce high-velocity, quasineutral pulsed plasma jets using the electromagnetic Lorentz force. Variations of such devices, including the quasi-stationary plasma accelerator (QSPA) \cite{Safronov2001,Belan1996,Chebotarev1996}, are capable of delivering energy fluxes to candidate materials that replicate ELM conditions in the laboratory. At Stanford University, a plasma gun was built for this purpose as an extension of the classic Marshall configuration \cite{Marshall1960}. The design, as detailed in figure~\ref{fig:theory}, features a 26 cm long coaxial accelerator volume with 0.5 cm diameter stainless steel rod anodes and a central copper cathode.

The production of plasma jets in accelerator experiments requires the expansion and eventual ionization of neutral gas in vacuum conditions. At the Stanford facility, neutral gas is supplied to the accelerator volume to initiate a discharge by a fast rise-rate, variable mass-bit gas puff valve, detailed in \cite{Loebner2015Valve}. A vacuum pressure of $10^{-7}$ Torr is maintained between successive firing events to maintain consistent and reliable operation of the valve and accelerator during the breakdown process. Energy is supplied to break down and accelerate the neutral gas by connecting a 56 $\mu$F capacitor bank across the electrodes, charged to voltages ranging from 5 to 12 kV.  As the neutral gas flows into the breech, it is ionized and accelerated out of the gun volume in a quasi-steady manner. The acceleration process is a result of a strong axial Lorentz (J$\times$B) force, created by the radial current flow between the concentric electrodes and the induced azimuthal magnetic field. The plasma flow can be succinctly described using the deflagration branch of the Rankine-Hugniot theory \cite{Loebner2015,Cheng1970,Loebner2016} and forms stable jets for timescales over which current and neutral gas can be supplied.

\subsection{Flow properties}

A major advantage of plasma guns is their inherent ability to reliably and repeatably expose material targets to extreme heat loads. However, detailed characterization of the resulting plasma plume is necessary to determine their suitability in simulating ELMs. At the Stanford facility, a suite of diagnostics have been employed to uncover the underlying plasma properties of the plume that interacts with material surfaces. Namely, spectroscopic line broadening has been employed to determine both plasma size and density \cite{Underwood2018}, time-of-flight to measure velocity \cite{Loebner2015, Underwood2018}, and distributed probes/detailed numerical simulations to quantify both magnetic field and plasma temperature \cite{Subramaniam2018,Poehlmann2010}. A survey of these results are included in table~\ref{tab:prop} along with measurements from similar facilities. Other experiments are also included that incorporate additional stages to the conventional accelerator geometry such as a CUSP trap and an extended tube drift \cite{Arkhipov1996}.

\Table{\label{tab:prop}Summary of plasma properties and performance characteristics of various gun facilities \cite{Federici2001,Safronov2001}.}
%\begin{indented}
%\item[]\begin{tabular}{@{} llllll}
\br
\br
&Jet Diam.& Pulse Len.& \hspace{4.5mm}Density&B-field& Ion Energy\\
Device & (cm) & ($\mu$s)& \hspace{4.5mm}(m$^{-3}$)  & (T) & (keV) \\
\mr
MK-200UG$^{\rm b}$  &    6.5    &    40-50         &   \hspace{7.6mm}$2 \times  10^{21}$     &     2       &  1.5\\
MK-200CUSP$^{\rm c}$      &    0.5   &  15-20   &  \hspace{0.55mm}(1.5-2$)\times10^{22}$      &   2-3      &   0.8\\
QSPA$^{\rm d}$                   &    5\lineup       &   250-600              &  \hspace{3.58mm}~$\sim1\times10^{22}$                       &   0-1         &   0.1\\
Kh-50   &  4       &   200                      &   \hspace{2.88mm}(2-8$)\times10^{21}$       & 0-2     & 0.3 \\
PLADIS                &   2       &        80\lineup-500           &   \hspace{10.85mm}---                                    &   ---         & 0.1\\
VIKA                     &   6     &    90-360                  & \hspace{4.9mm}$>1 \times 10^{22}$       & 0-3                     &0.2\\
Stanford Gun        &  1-2     &   10-20                  &    \hspace{5.7mm}10$^{22} $-10$^{23}$        &      0-1          &  0.1$^{\rm a}$\\
\br
\end{tabular}
\scriptsize
\item[] $^{\rm a}$ Ion energy calculated assuming hydrogen, $E = 0.5mV^2+1.5k_{b}T$, where $T\approx 25$ eV and $V \sim 100$ km/s from experiment/simulation \cite{Underwood2018,Subramaniam2018}.
\item[] $^{\rm b}$ Pulsed plasma gun coupled with long magnetic drift tube \cite{Arkhipov1996}.
\item[] $^{\rm c}$ Pulsed plasma gun with magnetic quadrupole \cite{Arkhipov1996}.
\item[] $^{\rm d}$ Remaining guns are all quasi-stationary accelerators with no added stages \cite{Belan1996, Chebotarev1996, Crawford1993, Loebner2016PMI}.
\end{indented}
\end{table}
The short duration ($<$1 ms), high plasma density ($>$10$^{22}$ m$^{-3}$), and low ion kinetic energy (100-300 eV) introduce uncertainty when extrapolating data from plasma guns to fusion conditions. Current projections of type I ELMs in next step fusion devices estimate plasma properties that originate from the pedestal step reference regime of ITER ($n_{ped} \approx 0.8 \times 10^{19}$ m$^{-3}$, $T_{ped} \approx 4$ keV) \cite{Loarte2007}. During ELMs, measurements indicate that the duration and rise time of power pulses are correlated with the ion transport time from the pedestal to the divertor target \cite{Loarte2003}. These measurements imply that the characteristic velocity by which ELM energy is transported to both the divertor target and chamber PFCs is controlled by the ion sound speed and thus the plasma properties originating from the pedestal region \cite{Pitts2006}.

Although similar heat fluxes are achievable with plasma jets from gun experiments, differences in flow properties lead to deviations in flow collisionality that substantially affect the rate of energy transfer to material surfaces. To illustrate this, consider the interaction of two separate plasmas with first wall materials, one with properties consistent with jets created by the Stanford plasma gun ($n \sim 10^{23}$ m$^{-3}$, $T \sim 25$ eV, $V \sim 100$ km/s) and one consistent with the pedestal region in ITER where type I ELMs originate ($n \sim 0.8 \times 10^{19}$ m$^{-3}$, $T \sim 4$ keV). To quantify collisionality, the ion-ion, $\nu_{s}^{i|i}$, and ion-electron, $\nu_{s}^{i|e}$, mean collision frequencies can be calculated \cite{Merritt2014} for each set of flow conditions using,
\begin{eqnarray}
      \nu_{s}^{i|i} =\cases{ 1.4 \times 10^{-13} n_{i} Z^{4} \ln \Lambda_{ii} \frac{ T^{-3/2}}{\mu^{1/2}}& $\displaystyle \frac{m_{i} V^{2}}{2 k_{b} T } \ll 1$  \label{eq:i|i}\\
      1.8\times 10^{-13} n_{i} Z^{4} \ln \Lambda_{ii} \frac{\epsilon^{-3/2}}{\mu^{1/2}} &  $\displaystyle \frac{m_{i} V^{2}}{2 k_{b} T } \gg 1$  \\},\\
       \nu_{s}^{i|e} = 1.6 \times 10^{-15} n_{e} Z^{2} \ln \Lambda_{ie} \frac{ T^{-3/2}}{\mu}, \label{eq:i|e}
\end{eqnarray}
where $n_{i}$ and $n_{e}$ are the ion and electron number densities in m$^{-3}$, $Z$ is the ion charge, $\mu = m_{i}/m_{p}$, $m_{i}$ and $m_{p}$ are the ion and proton masses respectively, $T$ = $T_{i}$ = $T_{e}$ is the plasma temperature in eV, and $\epsilon_{j} = 0.5 m_{j} V^{2}$ is the particle kinetic energy in eV. The expression in Eq.~(\ref{eq:i|i})-(\ref{eq:i|e}) considers only a single ion species and assumes that electron drag is thermally dominated ($m_{e} V^{2}/ 2 k_{b} T  \ll 1$). The Coulomb logarithm, $\ln \Lambda$, needed for evaluating the collision frequencies, can be expressed as,
 \begin{eqnarray}
\ln \Lambda_{ii} = \cases{43 - \ln \left[ \frac{2Z^{2}\times 10^{-3}}{\mu  (V/c)^{2}} \left(\frac{n_{e}}{T}  \right)^{1/2} \right] & $ \displaystyle  \frac{k_{b} T}{m_{i}} < V^{2} < \frac{k_{b} T}{m_{e}}$ \\
23 - \ln \left[ \frac{Z^{2}\times 10^{-3} }{T} \left(\frac{2 n_{i} Z^{2}}{T}  \right)^{1/2} \right] & else },\label{eq:ii}\\
\ln \Lambda_{ie} = \cases{ 23 - \ln \left( 10^{-3}n_{e}^{1/2} Z T^{-3/2} \right) & $T < 10 Z^{2}$ eV     \\
									  24 - \ln \left( 10^{-3} n_{e}^{1/2} T^{-1}  \right) &  $T >$ $10 Z^{2}$ eV    \\ } \label{eq:ie}.
\end{eqnarray}
For $\ln \Lambda_{ii}$, $V$ is taken as the bulk flow velocity for plasma guns to capture the significant axial kinetic energy present in the counter-streaming stagnation process and $c$ is the speed of light. The mean time between ion collisions, $\tau_{i}$, can be calculated using,
 \begin{eqnarray}
\tau_{i} = (  \nu_{s}^{i|e} + \nu_{s}^{i|i} )^{-1},\\
\tau_{i} =\left (  \nu_{s}^{i|e} + \frac{\nu_{s,S}^{i|i} \exp \left( \frac{2 k_{b} T}{m_{i} V^{2} }  \right)  + \nu_{s,F}^{i|i} \exp \left( \frac{m_{i} V^{2}}{2 k_{b} T}  \right)}{   \exp \left( \frac{2 k_{b} T}{m_{i} V^{2} }  \right)  +  \exp \left( \frac{m_{i} V^{2}}{2 k_{b} T}  \right)}  \right)^{-1},
\end{eqnarray}
where $\nu_{s,S}^{i|i}$ and $\nu_{s,F}^{i|i}$ are the slow ($m_{i} V^{2}/ 2 k_{b} T  \ll 1$) and fast limits of ($m_{i} V^{2}/ 2 k_{b} T  \gg1$)  of Eq.~(\ref{eq:i|i}) respectively \cite{Messer2013}.

\begin{figure*}[bp]
\centering
\includegraphics[width=0.85\textwidth]{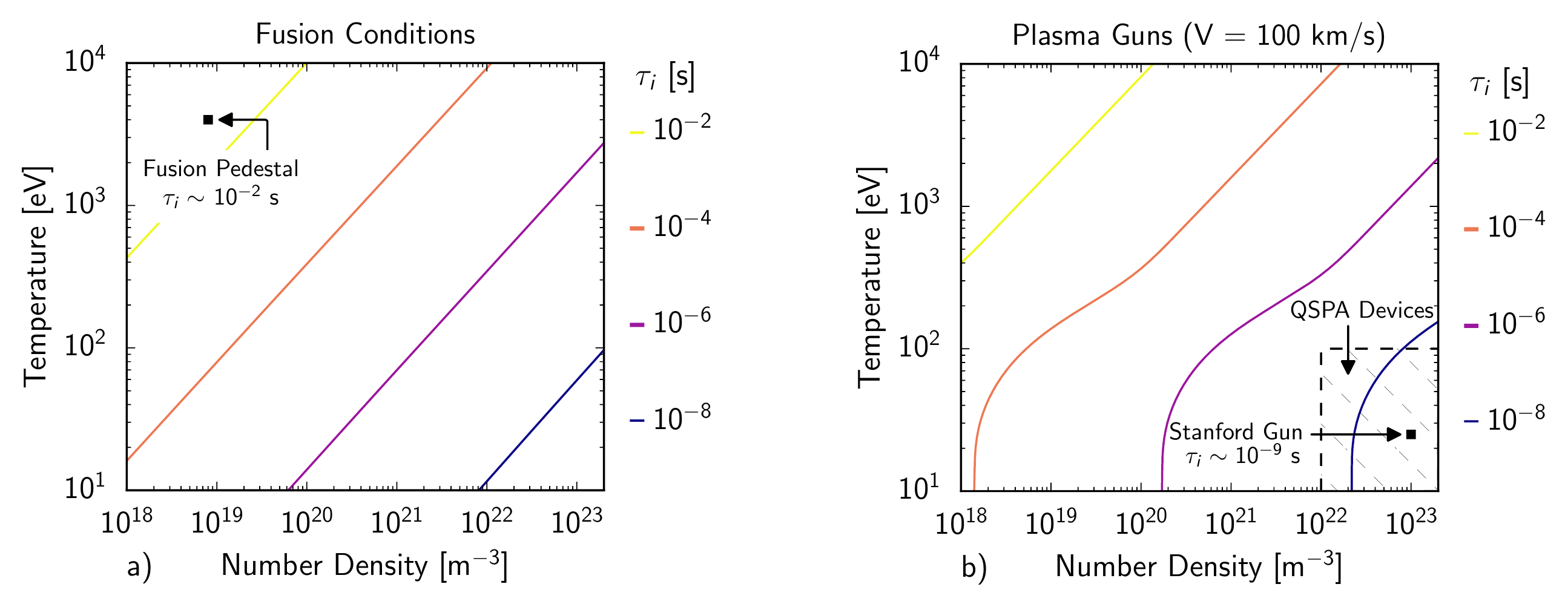}
\caption{Plot of mean ion collision time, $\tau_{i}$, as a function of both temperature and density for representative conditions in (a) fusion ELMs and (b) plasma guns. In general, plasma guns produce streams which are denser ($>$10$^{22}$ m$^{-3}$ typically) and colder ($<100$ eV typically) than fusion conditions. This leads to significant collisional effects in stagnation experiments. }
\label{fig:tau}
\end{figure*}

Results detailing $\tau_{i}$ for both flow conditions are included in figure~\ref{fig:tau} assuming a hydrogen ion stream ($Z=1$). The increased density and lower temperature in plasma gun experiments leads to collisional times of $\tau_{i} \sim 10^{-9}$ s. Given the jet lifetime, $\tau_{\textnormal{\scriptsize jet}}$, for ELM simulation ranges from 10 $\mu$s to 100 $\mu$s in gun devices, $\tau_{\textnormal{\scriptsize jet}} \gg \tau_{i}$ and thus collisional effects must be considered. In the pedestal region however, much longer collisional times of $\tau_{i} \sim 10^{-2}$ s are expected. Given that ELM durations in next step devices are expected to range from 0.1-1.0 ms \cite{Federici2001}, disturbances are expected to follow the predominant collisionless regime of fusion systems. As indicated in figure~\ref{fig:tau}(a), even if significant compression takes place during the transport of ELM energy from the pedestal to the divertor, the collisional timescale for gun devices will still significantly overestimate fusion conditions.

As $\tau_{\textnormal{\scriptsize jet}} \gg \tau_{i}$ , the result indicates that a selective shock shielding layer will form in front of the stagnation targets due to flow collisionality, as illustrated in figure~\ref{fig:theory}(b). This shock shielding effect is unique to the conditions of plasma gun flows and acts to reduce both the heat flux and particle kinetic energy reaching the wall. In the proceeding sections, we provide experimental proof through direct visualizations that shock layers form in gun experiments over timescales that precede vapor ablation formation. In addition, we use numerical simulations to quantify the effect this shielding layer has on both the heat flux and particle velocities reaching the wall. These results will help to connect experimental measurements made in plasma guns, where characteristic heat fluxes are typically estimated using plume properties upstream of shock layers, to the fusion conditions that motivate them.

\section{Shock Formation}

This section outlines the computational and experimental framework that underlie the simulations presented in this paper. Presented first are the magnetohydrodynamic (MHD) governing equations employed to describe the production, acceleration, and eventual stagnation of highly collisional plasma jets formed in gun devices. This is followed by a brief description of the numerical methodology used to spatially discretize and time-integrate the MHD system. Finally, time-resolved experimental schlieren images will be presented to show the shock shielding that forms in collisional QSPA devices and further validate the numerical model. This framework will be used to investigate the effect of collisionality on energy and particle transport to material interfaces.

\subsection{Computational model}
\subsubsection{Governing equations} \label{sec:GE}

The MHD governing equations are a `single-fluid' description of a quasi-neutral high density plasma \cite{Goldston1995,Inan2010}. The high pressure ($>$1 atm) operational regime, underlying breakdown and ionization in plasma guns, ensures short energy transfer mean free paths and thus rapid temperature equilibration between all constituent species \cite{Subramaniam2018v2}. Hence, the local thermodynamic equilibrium (LTE) approximation is used to define plasma composition and its thermodynamic properties. The high collisionality between the constituent plasma species also justifies their treatment using a single bulk velocity. Since the plasma is predominantly quasi-neutral and has a high ionization fraction, the small-scale sheaths formed around material surfaces are not resolved within this formulation. Furthermore, since the timescale of plasma-surface interactions precedes vapor ablation, the physics of ablation, namely material removal from the target surface and material injection into the plasma, is not modeled. The large Reynolds number and fast plasma transient allow for the viscous and conductive heat transfer effects to be ignored within the bulk of the plasma jet.  

Using the aforementioned assumptions, the MHD equation system is derived by coupling the Navier-Stokes and Maxwell equations \cite{Bittencourt2004,Boulos2014}. The Navier-Stokes equations describe the fluid dynamic behavior of the plasma, influenced by the Lorentz force and Joule heating source terms that accelerate the discharge. The source terms are in turn computed by solving the magnetic vector induction equation, derived from the Maxwell’s equations that describe the electromagnetic fields and current density sustained by the discharge. The entire system is closed by an equation of state and Ohm's law. The resistive MHD equations in conservative form are written as,
\begin{eqnarray}
\fl \frac{\partial \rho}{\partial t} + \nabla \cdot \left( \rho \bm{V} \right) = 0, \label{eq:cont}\\
\fl \frac{\partial}{\partial t} \left( \rho \bm{V} \right) + \nabla \cdot \left( \rho \bm{V} \bm{V} \right) - \nabla \left( P + \frac{\bm{B} \cdot \bm{B} }{2 \mu_{0}} \right) = 0,\\
\fl \frac{\partial \bm{B}}{\partial t}  + \nabla \cdot \left( \bm{V} \bm{B} - \bm{V} \bm{B} \right) = - \nabla \times \left( \eta \frac{\nabla \times \bm{B}}{\mu_{0}} \right),\\
\fl \frac{\partial}{\partial t} \left( \rho e_{t} + \frac{ \bm{B} \cdot \bm{B}}{2\mu_{0}} \right) + \nabla \cdot \left(  \left[ \rho e_{t} + \frac{\bm{B} \cdot \bm{B}}{\mu_{0}} + P \right] \bm{V} - \frac{\bm{V} \cdot \bm{B}}{\mu_{0}} \bm{B} \right) = - \frac{1}{\mu_{0}} \nabla \cdot \left[ \eta \left( \nabla \times \bm{B} \right) \times \bm{B} \right),\\
\fl \nabla \cdot \bm{B} = 0.\label{eq:divB}
\end{eqnarray}
Here $\rho$, $P$, $\bm{V}$, and $\bm{B}$ represent the plasma density, pressure, velocity and magnetic induction vector respectively. Additionally,  $\eta$ represents the plasma resistivity while $e_{t}$  represents the specific total energy, composed of the specific internal energy  $e$ and the specific kinetic energy $|(\bm{V}\cdot\bm{V})/2|$, that is $e_{t} = e + |(\bm{V}\cdot\bm{V})/2|$. The plasma resistivity is a function of the plasma pressure and temperature.  A tabulation based on the Spitzer resistivity model \cite{Boulos2014} is used to calculate the plasma resistivity as a function of the temperature and pressure of the plasma. The Spitzer resistivity \cite{Spitzer1953} is given by,
\begin{eqnarray}
\eta = \frac{ \left( 5e - 5 \right) \ln \Lambda}{T^{5/2}}. \label{eq:eta}
\end{eqnarray}
The tabulation uses the hydrogen resistivity data from [6] to back out the parameter $\ln \Lambda$, the coulomb logarithm. The equation of state is used to close the system by relating the internal energy, pressure, density and temperature.  These relations are given by,
\begin{eqnarray}
e(T) &= \frac{RT}{\gamma - 1},\\
P &= \rho R T, \label{eq:P}
\end{eqnarray}
where $\gamma$ is the ratio of specific heats at constant pressure and volume respectively. A consistent description of the dense thermal plasma in coaxial plasma accelerators is provided using Eq.~\ref{eq:cont}-\ref{eq:divB} with the closure relations given by Eq.~\ref{eq:eta}-\ref{eq:P}.

\subsubsection{Numerical methodology}
For the sake of brevity, presented here are certain salient features of the numerical model. A detailed description of the MHD computational tool can be found in \cite{Subramaniam2018, Subramaniam2017}, where it is used to study mode transition in plasma guns. The equations are spatially discretized using a cell-centered finite volume scheme on generalized unstructured grids \cite{Subramaniam2017}. The convective fluxes are obtained using approximate Riemann solvers such as the Harten-Lax-VanLeer (HLL) \cite{Toro2009} scheme or the local Lax-Friedrichs \cite{Leveque2002} method. The diffusive fluxes are obtained by gradient reconstruction followed by cell averaging. The spatially discretized ODE system is time-integrated using a fully-implicit backward Euler method. The algebraic system arising as a result of the fully-implicit discretization is solved using a multi-dimensional Newton method. The flux Jacobians that constitute the derivative of the objective function are derived analytically for a given flux scheme on generalized unstructured grids. The sparse linear systems generated during the Newton iterations are solved using the Generalized Minimum Residual (GMRES) method implemented in the suite of Krylov Subspace solvers available in the Portable and Extensible Toolkit for Scientific Computing (PETSc) library \cite{Balay2010}.

\subsubsection{Plasma-Vacuum interface tracking scheme}
One of the main challenges associated with using the MHD governing equations to simulate the deflagration mode arises due to the presence of a hard-vacuum initial condition. Within the expanding jet, the Knudsen number (ratio of particle mean free path to the gradient length scale of the plasma) increases as one moves from the high density core of the jet to its rarefied peripheries. Hence, the continuum based MHD equations lose their validity towards the regions of the jet that interact with the vacuum background. Prior efforts \cite{Subramaniam2017, Sitaraman2014} to simulate the deflagration employ a fictitious low density background gas to mimic vacuum. However, simulations performed using this method display an artificial shock at the plasma vacuum front. Not only is the shock inconsistent with the physics of free expansion, a process characterized by a smooth decrease in temperature mediated by a single expansion wave, the low background densities generate deflagration plumes that are qualitatively different from those observed in experiments. 

    To overcome this issue, a plasma-vacuum interface tracking scheme has been developed \cite{Subramaniam2018Vacuum}. The method works by identifying the high density regions of the jet where the MHD formulation is valid and tracking this domain of validity through the entire course of the simulation. The domain is further defined as the collection of cells (plasma cells) in the computational mesh that have a density higher than a predetermined threshold value. Once identified, the MHD computational framework is used to model the jet dynamics exclusively within the plasma cells. The collection of plasma cells increases as the deflagration jet expands to fill up the initially evacuated domain with a quasi-neutral plasma. Simulations performed using this framework indicate a deflagration jet spatial profile devoid of the artificial shock observed in the low background density simulations. Furthermore, the temperature smoothly varies from high values in the core of the jet to comparatively lower values at its peripheries, a result that is consistent with the physics of vacuum expansion and qualitatively in agreement with experimental observations.

\subsection{Experimental Visualization}

In support of the numerical investigation, a z-type schlieren configuration \cite{Settles2012, LoebnerThesis} was used to visualize the stagnation of collisional plasma jets on material targets. This diagnostic is a  form of refractometry that is well suited to resolve shocks and flow features in stagnation experiments. In plasma systems in particular, gradients in plasma density and sharp flow boundaries are easily captured as gradients in the measured image intensity with tunable sensitivity.

\begin{figure*}[bp]
\centering
\includegraphics[width=0.55\textwidth]{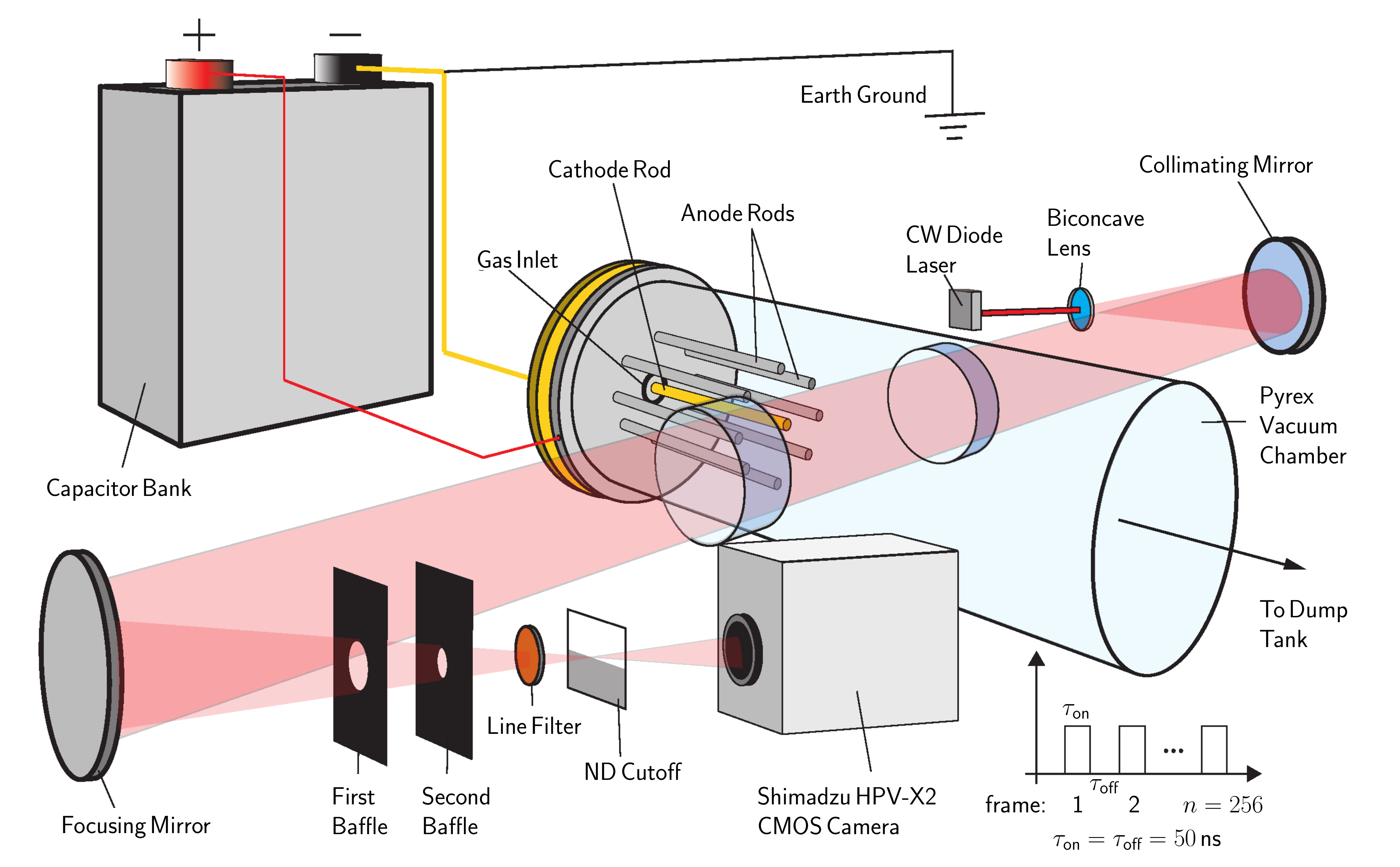}
\caption{Experimental setup used to visualize shock formation in the Stanford plasma gun using the z-type Schlieren imaging technique. Specific features of the optical setup include $f/4$, 60 cm focal length mirrors, a high-speed (10 MHz) camera, a laser backlight, and a `sooted' slide cutoff to limit diffraction effects. Diagram adapted from \cite{LoebnerThesis}.}
\label{fig:schlieren}
\end{figure*}

A detailed diagram of the optical setup used in the visualization experiment is shown in figure~\ref{fig:schlieren}. The traditional z-type schlieren configuration was modified to overcome the challenges presented by a pulsed plasma system. Namely, a laser was used in place of more conventional incoherent illumation sources to overcome the self-emission present in gun devices. A `sooted' slide was also used as the optical cutoff the adjust sensitivity and limit diffraction effects induced by the laser backlight source. The diagnostic was imaged with a Shimadzu HPV-X2 high-speed camera to capture up to 256 individual 50 ns exposures. This optical system permits the visualization of flow stagnation over both spatial and temporal scales necessary to capture shock growth and isolate it from vapor dynamics that have been studied over longer timescales \cite{Belan1996}.

\begin{figure*}[bp]
\centering
\includegraphics[width=0.9\textwidth]{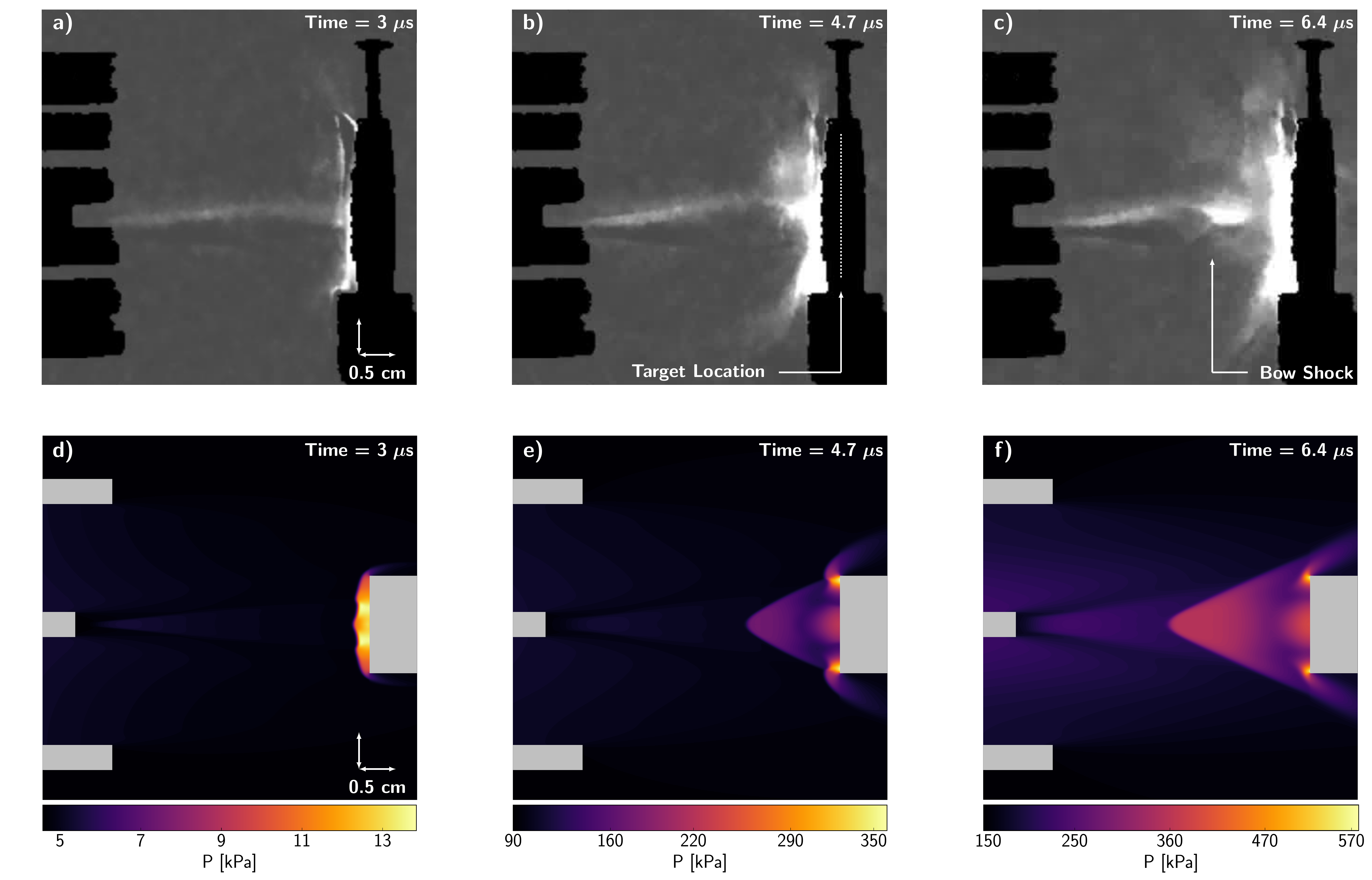}
\caption{Schlieren images, (a) - (c), and accompanying simulations, (d) - (f), of the production and subsequent stagnation of a plasma jet onto a 2 cm diameter tungsten target placed 3.6 cm downstream of the Stanford accelerator charged to 1.4 kJ. Both simulations and experiments show an extended bow shock forms on the front of the target before vapor shielding effects can establish.}
\label{fig:stagnation}
\end{figure*}

Schlieren images and accompanying simulations of plasma jet stagnation onto a target downstream of the accelerator volume is shown in figure~\ref{fig:stagnation}. In the experiment, the plasma jet was produced by the Stanford gun facility, depicted in figure~\ref{fig:theory}, charged to 1.4 kJ and stagnated onto a 2 cm diameter tungsten target placed 3.6 cm downstream of the gun volume. Complementary MHD simulations, performed on the same geometry, capture the production, acceleration, and stagnation of the plasma jet onto a material surface. Experimental current traces, shown in figure~\ref{fig:current} for a variety of charging energies, were imposed in the simulation to capture the LRC circuit response in the gun device. Figure~\ref{fig:stagnation}(d)-(f) show the simulated spatial distribution of fluid pressure in the vicinity of the target surface at the same times as the schlieren images in figure~\ref{fig:stagnation}(a)-(c). At 3 $\mu$s, the supersonic (80 km/s) jet impacts the target generating a thin normal shock with post-shock temperatures of approximately $40$ eV. However, as the plasma jet continues to interact with the target, sustained pressure deposition on the surface leads to a transition in the shock structure from a normal shock at 3 $\mu$s to a detached bow shock at 4.7 $\mu$s. From 4.7 $\mu$s to 6 $\mu$s, the bow shock continues to grow in size while maintaining its characteristic structure. Local pressure and temperature hot spots are obtained at the corners of the target as the maximal pre-shock free stream velocities are observed at these regions.

Both experiment and simulations indicate that an extended bow shock forms on the front of the material target and grows over the lifetime of the stagnating jet. Uncertainty in the initial gas distribution that is puffed into the gun volume contribute to discrepancies in the exact standoff distance between experiments and simulations. However, the presence of an abrupt pressure rise, decrease in axial velocities to near-stagnation values, and sharp deviation in flow streamlines are all conclusive evidence that the structure in the target vicinity is a shock generated by the high collisionality within the plasma jet. The remaining focus of the paper is to quantify the effect of self-shielding in plasma gun experiments and how existing results can be interpreted when extrapolating to fusion conditions.

\section{Shock Shielding}

\begin{figure*}[bp]
\begin{minipage}[H]{0.5\textwidth}
  \centering
\includegraphics[width=0.8\textwidth]{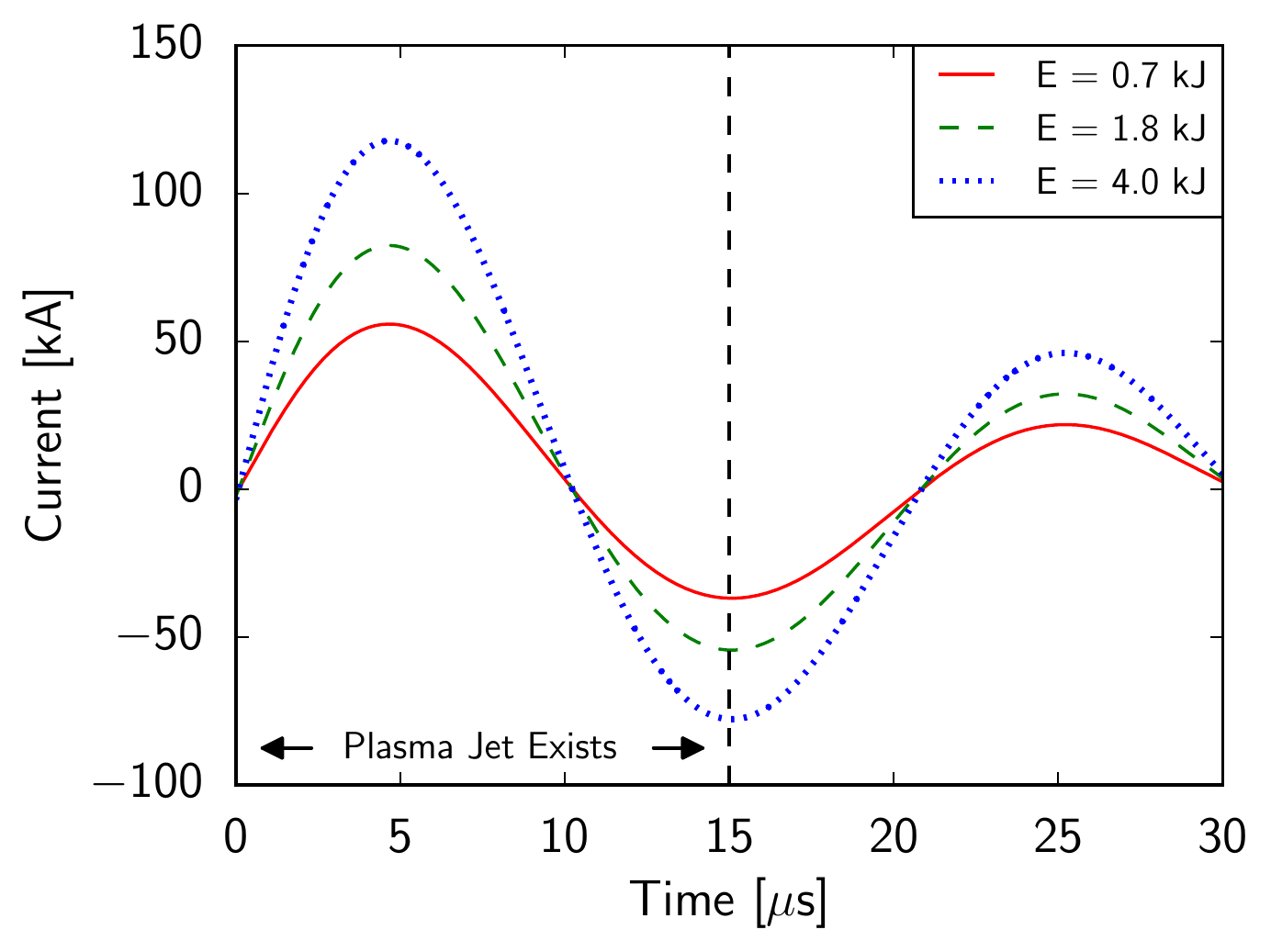}
    \caption{Experimentally measured current profiles of the plasma guns at a variety of charging energy used to simulate the LRC circuit response of the gun device. For all simulation results, only the first 15 $\mu$s of the current trace is analyzed where a plasma jet exists.}
    \label{fig:current}
  \end{minipage}
  \hfill
  \begin{minipage}[H]{0.5\textwidth}
  \centering
\includegraphics[width=0.55\textwidth]{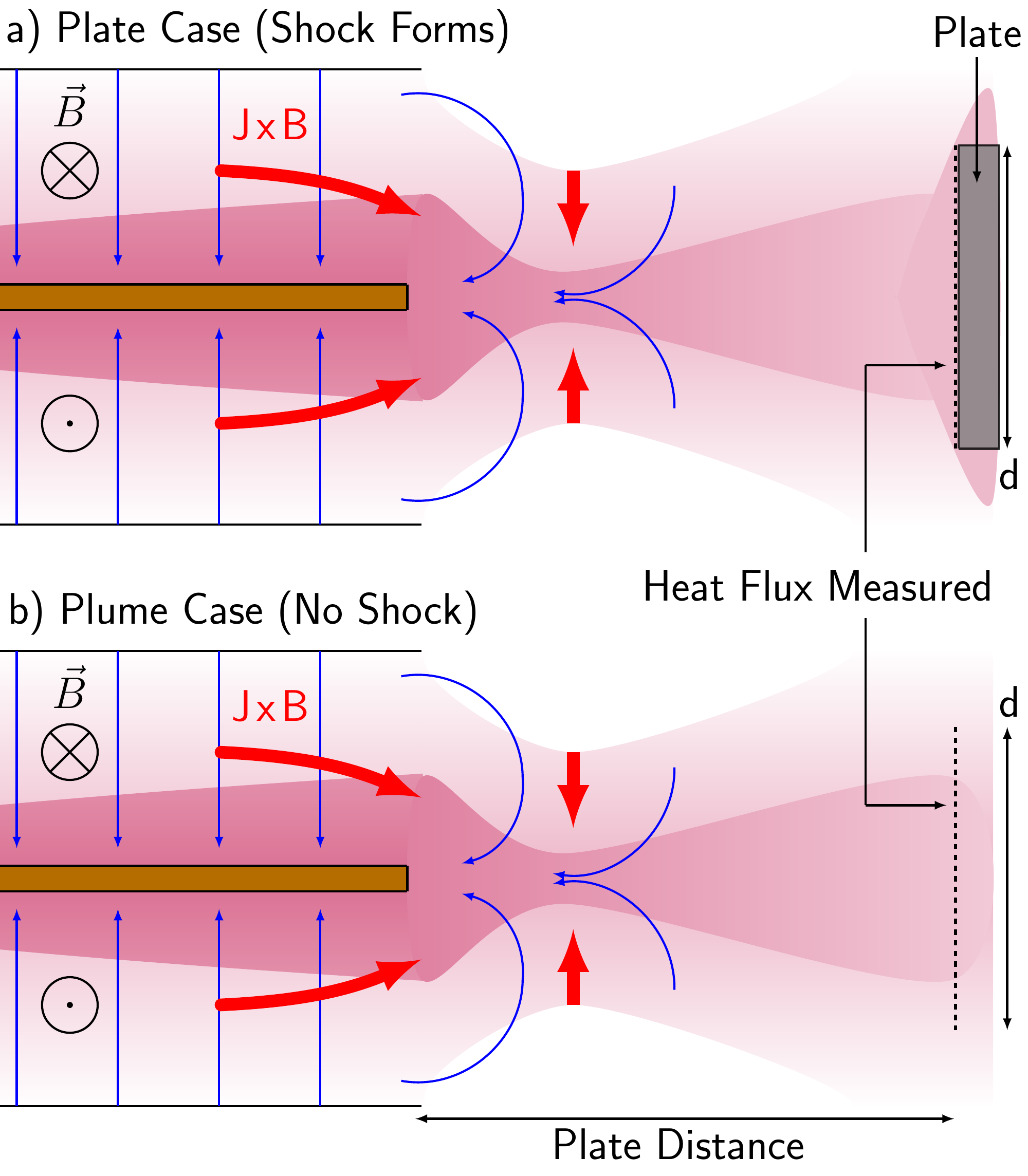}
    \caption{Diagram detailing plate and plume cases used to isolate the effect flow collisionality has on particle energy transport in the vicinity of a material interface. The plume case simulates the same jet structure but estimates the collisionless regime expected in fusion conditions by not allowing energy redistribution due to shock formation. Both cases track properties over a surface area equal to the plate area for a $d = 2$ cm circular target.}
    \label{fig:plumePlateCase}
  \end{minipage}
\end{figure*}

This section investigates the quantitative effect that shock shielding has on both the incident heat flux and the particle energy spectrum over a span of conditions accessible by gun experiments. Specifically, the variation in the flow shielding effect as a function of the incident particle flux is explored by varying the gun charging energy, detailed in figure~\ref{fig:current}, and the downstream distance to the target. For the collisional regime of plasma guns, shocks act to convert the axially directed kinetic energy of the flow to thermal energy and further, reduce the transport velocity of the incoming particle flux to material interfaces. To capture this effect and establish a reference point to fusion conditions, two cases were included for each of the conditions considered. The first case, hereon called the plate case (figure~\ref{fig:plumePlateCase}(a)), simulates the stagnation of a plasma jet generated by the Stanford plasma gun and tracks the flow properties in the vicinity of the target.  The second case, the plume case (figure~\ref{fig:plumePlateCase}(b)), is without a target and allows the plasma jet to freely expand into vacuum. These simulations are used to isolate the role that collisionality plays on the redistribution of energy in the vicinity of material targets.

In both the plate (with stagnation shock) and plume (no shock) cases, the axially directed particle heat flux in the plasma is tracked using \cite{Sutton1965},
 \begin{eqnarray}
\Gamma_{\epsilon,z} = V_{z} \left[ \frac{5}{2} p + \frac{1}{2} \rho V_{z}^{2} \right] + \sum_{s} \int c_{z} \left( \frac{m_{s}}{2} c^{2} \right) f_{s} dc, \label{eq:energyFlux}
\end{eqnarray}
where $p$, $\rho$, and $V_{z}$ are variables calculated in the MHD formulation while the last term in Eq.~(\ref{eq:energyFlux}) represents heat conduction in the flow around the plate. Comprehensive modeling of this term requires not only resolving the shock structure but also capturing the thermal boundary layer, electrical sheath, ablation/vapor dynamics, and radiation transport effects. Instead of detailed modeling of the material processes as in previous studies \cite{Hassanein1999}, this work focuses on modeling the effect collisionality has on the production and subsequent stagnation of plasma jets in gun devices.  

Detailed kinetic simulations of heat transport within plasma indicate that the actual conduction heat flux vector is a fraction of the free streaming limit, $q_{F}$. $q_{F}$ is calculated assuming that the total thermal energy content, $n k_{b}T$, of the plasma volume streams in the gradient direction with the characteristic thermal velocity \cite{Bell1985}. Using this limit, the relative effect the shock has on altering the available heat flux in the vicinity of the plate can be calculated. Assuming a Maxwellian velocity distribution for a hydrogen stream and further assuming the ambipolar particle flux criterion enforced by the sheath, $\Gamma_{i} = \Gamma_{e}$, the fluid heat flux presented in Eq.~(\ref{eq:energyFlux}) becomes,
 \begin{eqnarray}
\Gamma_{\epsilon,z} = V_{z} \left[ \frac{5}{2} p + \frac{1}{2} \rho V_{z}^{2} \right] + n k_{b} T\bar{c}.
\end{eqnarray}
In the expression for $\Gamma_{\epsilon,z}$,  $\bar{c} = \sqrt{8 k T/ \pi m_{p}}$ represents the average thermal speed of the hydrogen ion species, with mass $m_{p}$, that controls the thermal flux transport. To explain discrepancies in heat flux induced by collisionality, a control volume analysis is also presented to capture the transfer of energy within the shock volume. In all cases, the analysis is framed to fusion applications to understand how experimental measurements made with collisional plasma guns can be interpreted.

\subsection{Energy Scaling}

\begin{figure*}[tp]
\centering
\includegraphics[width=1.0\textwidth]{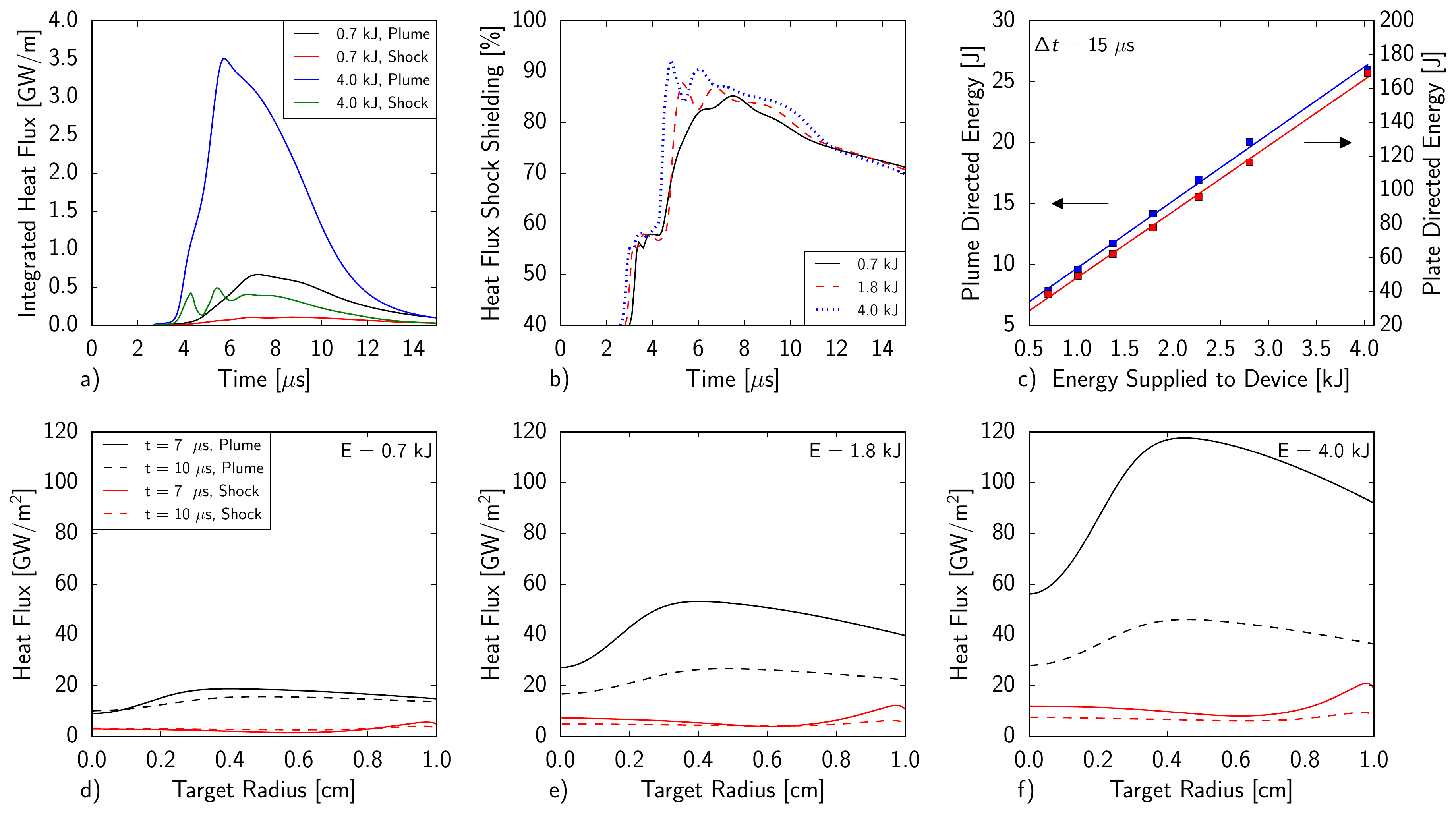}
\caption{Plots detailing characteristic flow properties recorded 4 cm downstream of the accelerator volume for both the plate and plume cases over a variety of charging energies.  (a) Details the spatially average heat flux over a region equal to the target diameter (2 cm) in the free stream limit while (b) quantifies the shielding effect the shock plays over time. (c) Integrates the heat flux over the jet lifetime ($\Delta t = 15$ $\mu$s) and target area and (d) - (f) plot the heat flux as a function of radius at select times.}
\label{fig:energyVar}
\end{figure*}

Tabulated flow properties for both the plume and plate cases featuring a variety of gun charging energies is detailed in figure~\ref{fig:energyVar}. In each case, the heat flux was evaluated using properties recorded 4 cm axially downstream of the accelerator, consistent with the target location. The current used in each charging condition was imposed based on experimental measurements shown in figure~\ref{fig:current}. Results in figure~\ref{fig:energyVar} indicate a strong self-shielding effect induced by the formation of a collisional shock in the plate cases. The amplitude of this effect is shown in figure~\ref{fig:energyVar}(a)-(b) where the shielding due to the formation of the shock, Shielding = $1- \Gamma_{\epsilon,\textnormal{\scriptsize shock}}/\Gamma_{\epsilon,\textnormal{\scriptsize plume}}$, is found to be up to 90\% of the incident plume heat flux.

Several oscillations in the free stream shock shielding amplitude are observed in figure~\ref{fig:energyVar}. These oscillations are also reflected in the integrated heat fluxes, detailed in figure~\ref{fig:energyVar}(a), for the plate and plume cases respectively. Specifically the plume cases in figure~\ref{fig:energyVar}(a) feature a smooth time variation with a single maximum concurrent with the maximum in the driving current pulse. However, the integrated flux for the plate cases show two individual peaks. For the 4.0 kJ case specifically, the first peak is observed at approximately $4.2$ $\mu$s and the second at approximately $5.7$ $\mu$s.

The spatial distribution of both the pressure and density in the vicinity of the target provide insight into the shock dynamics responsible for the heat flux oscillations. The plasma jet impacts the target at $3$ $\mu$s and forms a thin normal shock. This corresponds to the first spike in the shock shielding percentage plot, figure~\ref{fig:energyVar}(b). From 3 to 4.2 $\mu$s, the shock structure grows, transitioning from a normal to a detached bow shock that contains two regions of maximum pressure. The first of these is the bow shock tip, situated on-axis where the flow turn angle is minimal (locally normal shock) and the second is the plate surface where the post-shock flow stagnates. Between 3 and 4.2 $\mu$s, the plate surface consistently observes higher pressures and hence the integrated heat flux during this period continues to rise. At 4.2 $\mu$s, due to the rising free stream velocities, the location of peak pressure transitions to the bow shock tip. This causes the shock standoff distance to abruptly start increasing between 4.2 to 4.9 $\mu$s. During this phase, since the pressure energy is being effectively distributed over a larger volume, the plate pressure temporarily decreases, leading to a dip in the integrated heat flux. This corresponds to the second spike in figure~\ref{fig:energyVar}(b). At 4.9 $\mu$s however, the bow shock once again attains a stable standoff distance and the plate pressure and heat flux continue to rise until approximately 5.7 $\mu$s when the incoming free stream heat flux is maximum. Following the peak, the driving current pulse starts to fall, decreasing the available free stream energy and consequently the heat flux transferred through the shock shows a sharp decrease. This corresponds to the third peak in figure~\ref{fig:energyVar}(b). Following this, the bow shock stand off distance falls to reach a new equilibrium and the heat flux continues to smoothly decrease to zero.

In addition to actively reducing the applied heat flux to material interfaces, figure~\ref{fig:energyVar} also indicates that shocks alter the dynamics and structure of the heating profile. This effect is shown in figure~\ref{fig:energyVar}(d)-(f) where the heating profile for the plume and plate cases are presented at select times for varying energies. In each case, the shock acts to reduce the heat flux and redistribute its energy to peak near both the plate centerline and edge where maximum stagnation pressures are achieved. The observed shielding effect indicates that not only is the self-shielding of plasma guns substantial, but more importantly, estimates of both the plasma heat flux and energy densities based on upstream plume properties are substantially different than conditions that are actually imposed on material targets.

\subsection{Length Variation}

\begin{figure*}[bp]
\centering
\includegraphics[width=1\textwidth]{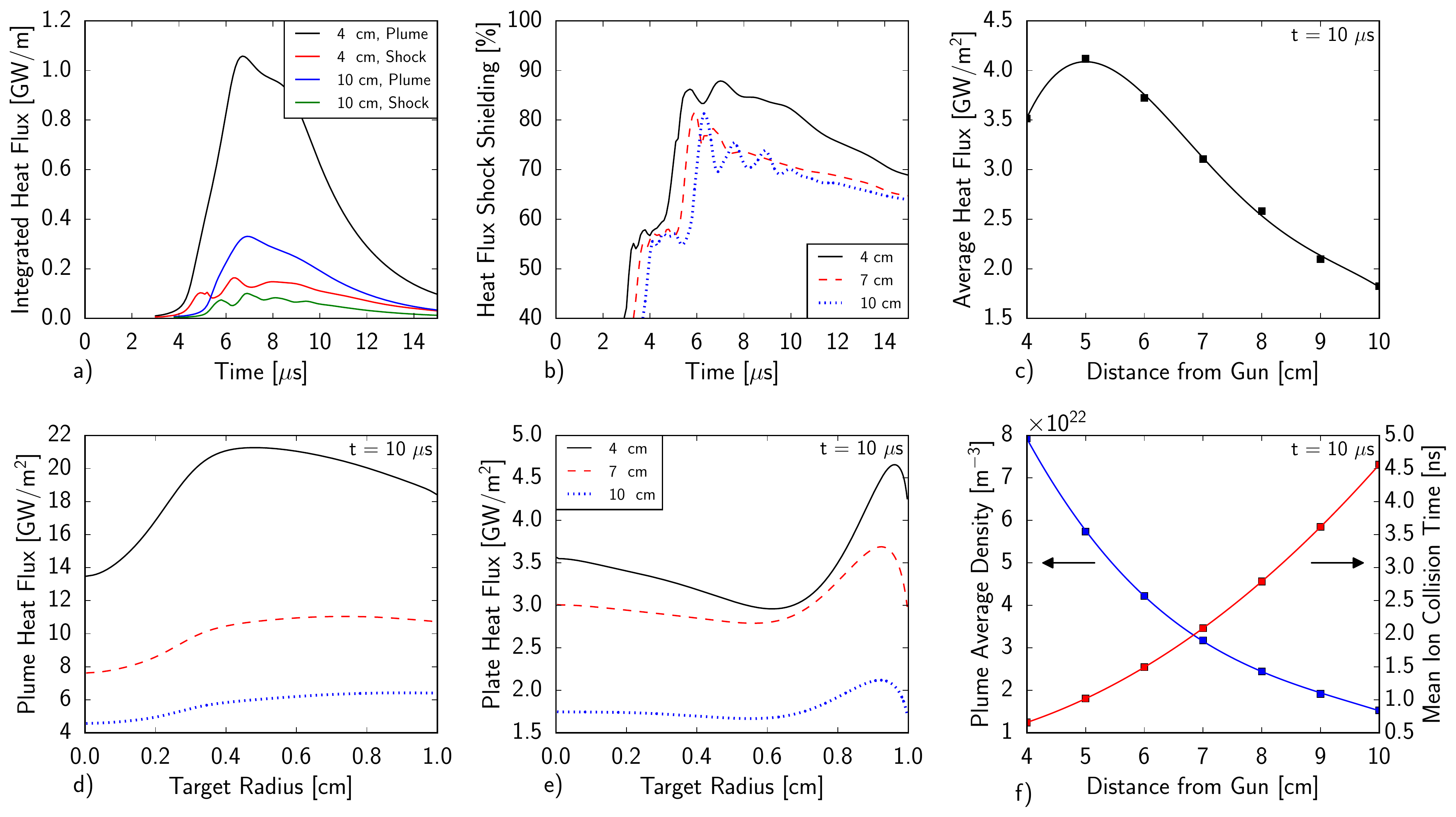}
\caption{Plots detailing characteristic flow properties recorded at various axial distances downstream of the accelerator volume for both the plate and plume cases with a charging energy of 1.4 kJ. (a) Details the spatially average plate heat flux over a region equal to the target diameter (2 cm) while (b) quantifies the shielding effect the shock plays over time. (c) Presents how the spatially average heat with the target position while (d) - (e) plot the heat flux as a function of radius for both the plate and plume cases at $t= 10$ $\mu$s. Finally(f) quantifies how the plume density drops with increasing axial distance due to radial plume expansion and the corresponding increase in collision time.}
\label{fig:lengthVar}
\end{figure*}

The results of the length variation for both the plate and plume cases are detailed in figure~\ref{fig:lengthVar}. In each case, a charging energy of 1.4 kJ was applied to drive the gun discharge and the flow properties were recorded at locations corresponding to the target location ranging from 4 cm to 10 cm downstream of the accelerator. The results in figure~\ref{fig:lengthVar} detail the temporal characteristics of the integrated and spatially averaged heat flux, shock shielding amplitude, and the plume properties as a function of the target location.

As with the energy scaling simulations detailed previously, the free stream shock shielding amplitude shows three peaks for a target distance of 4 cm. The first peak corresponds to the jet impact, the second to when the bow shock abruptly grows in axial extent thereby decreasing the pressure on the surface and the third peak to when the free stream velocity starts to fall. When the plate is moved to 10 cm downstream of the accelerator, two additional maxima are observed in the free stream velocity decay phase of the discharge as shown in figure~\ref{fig:lengthVar}(a)-(b). As was the case with the 4 cm simulations, these peaks are caused by a restructuring of the bow shock that forms on the target. Consequently, each shock stabilization event is associated with a dip in the surface heat flux and a peak in the effective shielding percentage as seen in figure~\ref{fig:lengthVar}(b).

The length variation studies detailed in figure~\ref{fig:lengthVar} further suggest that the heat flux self-shielding effect reduces in amplitude as the target is moved farther away from the gun. This trend is a consequence of the radial expansion of the jet where the average plasma density in the plume, shown in figure~\ref{fig:lengthVar}(f), reduces as it propagates axially. The farther downstream the stagnation target is placed, the less directed axial kinetic energy is available to be shielded by the subsequent shock that forms.

Although there is reduced shielding observed as the target is moved farther away from the accelerator, the available heat flux also goes down and reaches an optimal location where combined losses due to radial expansion and shock effects are minimized, as detailed in figure~\ref{fig:lengthVar}(c). This sets a limit on the regions over which a target can be placed and still receive type I ELM heat loading (1-10 GW/m$^{2}$). Beyond heat effects, accurate ELM simulation of material degradation with gun devices also requires consistency with the particle energy spectrum reaching the wall in fusion conditions. This places requirements on collisionality in gun devices which act to reduce the characteristic velocity of incoming particles fluxes. Results of the length variation studies indicate that while the average density in the plume drops as it propagates axially, the mean ion collision time is still much smaller than the jet lifetime $\tau_{\textnormal{\scriptsize jet}} = 15$ $\mu$s. This indicates that collisional effects are important through the accessible region where type I ELM simulation is possible with QSPA accelerators.

\subsection{Control Volume Analysis}

\begin{figure*}[tp]
\centering
\includegraphics[width=0.9\textwidth]{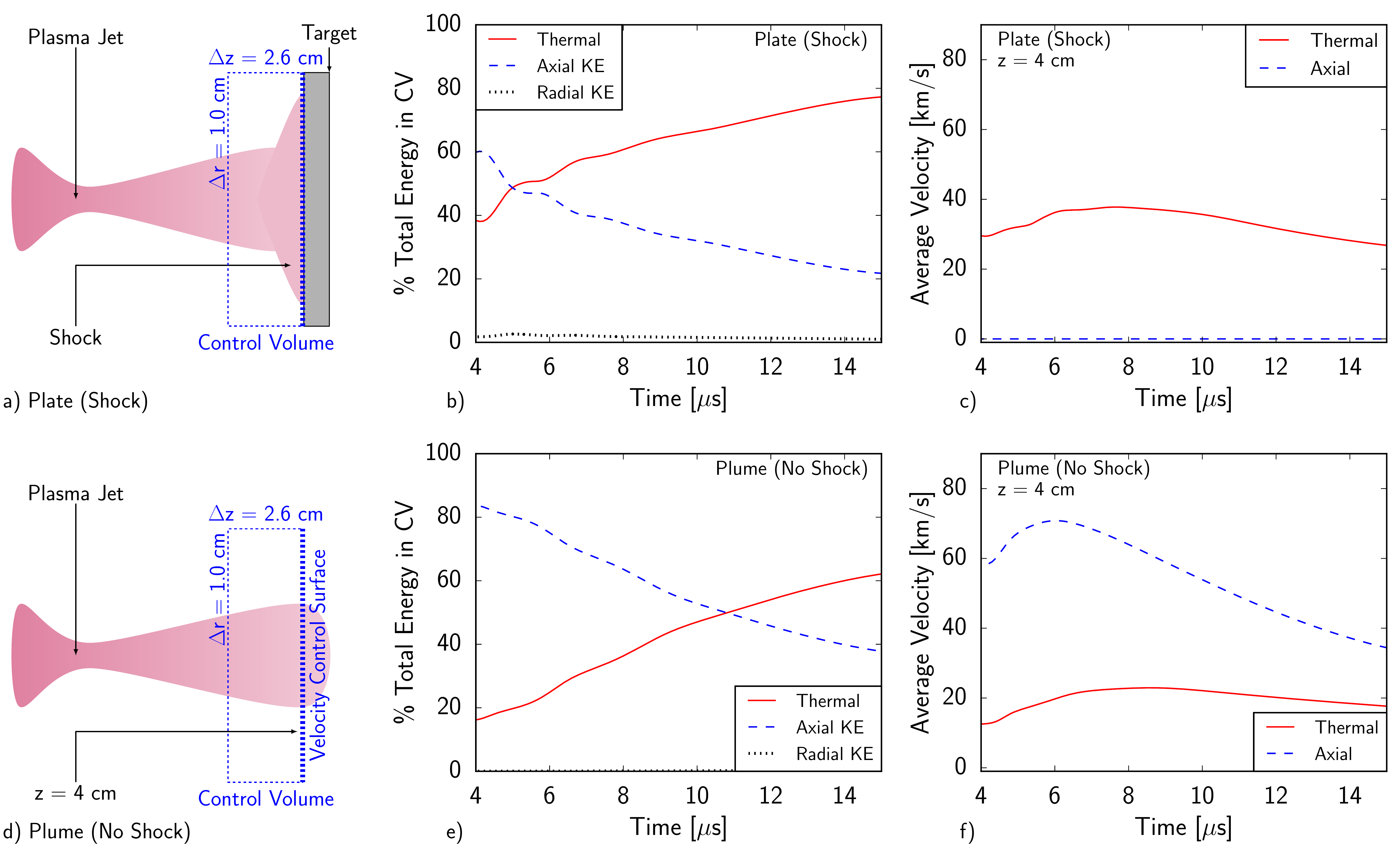}
\caption{Details of the control volume analysis performed for both the plate and plume cases with a charging energy of 1.4 kJ. In both cases, as depicted in (a) and (d), a control volume with an axial extent of 2.6 cm and a radial extent of 1 cm was placed 4 cm downstream of the accelerator to fully envelope the shock that forms on the stagnating target. Results of the control volume analysis are used to determine the distribution of energy after the shock and the change in velocity of the flow reaching the wall.}
\label{fig:CV}
\end{figure*}

A control volume (CV) analysis was performed for both the plate and plume cases to isolate how the shock redistributes energy within the plasma jet and ultimately changes the fluid velocity reaching material targets. In each case, a fixed control volume was placed 4 cm downstream of the accelerator to coincide with the target location and was defined to fully envelope the shock over the lifetime of the plasma jet. As depicted in figure~\ref{fig:CV}(a) and (d), this resulted in a CV with an axial extent of 2.6 cm and a radial extent of 1 cm. 

Results of the CV analysis for a gun charging energy of 1.4 kJ are presented in for plate conditions where a shock is generated in figure~\ref{fig:CV}(a)-(c) and the plume conditions in (d)-(f). In stagnation conditions captured in the plate case, the directed kinetic energy is redistributed as both radial kinetic energy and thermal energy in the vicinity of the plate. The plume conditions on the other hand retain both the thermal and directed contributions to the heat flux and feature time variations in the energy components only due to changes in the drive current responsible for the fluid acceleration. In the process of redistributing the energy, figure~\ref{fig:CV}(c) and (f) show how the bulk fluid and thermal velocities are changed with the addition of a shock. The fluid stagnation in a collisional fluid forces the axial bulk velocity to approach zero around the material interface and in doing so, reduces the energy transport rate via the thermal component of the velocity. This effective reduction in the transport rate of both particles and energy to the surface is responsible for the reduced heat fluxes observed in collisional fluids and overcomes the increased density and temperature found in stagnation conditions. Equally importantly for ELM experiments, figure~\ref{fig:CV} also indicates that collisional fluids significantly reduce the particle velocities and kinetic energies reaching material targets.

\section{Conclusion}

As research progresses toward next step fusion devices, loss of stable confinement continues to be a technical hurdle that must be overcome for reliable and prolonged operation. Experimental investigations into type I ELMs in particular are important to uncover the material degradation effects expected in operating fusion conditions. However, detailed characteristics of such events, with magnitudes consistent with ITER conditions, are not possible in existing tokamaks. 

Plasma guns have been used extensively to fill this void and simulate type I ELMs in the laboratory. They have distinct advantages over other techniques such as electron or laser beams where penetration into vapor layers are inconsistent with conditions expected in fusion reactors. This study however found that over timescales that precede this formation, an additional significant self-shielding due to flow collisionality is present in gun devices. Shock shielding effects cause up to 90\% of the free streaming heat flux to be shielded from the surface as the plasma stagnates against a material interface and creates a collisional shock. Coupled to this effect, the effective velocity of particles reaching the surface and thus the rate of energy transfer is reduced as the collisionality of the flow thermalizes the incoming free stream jet, eliminating the axially directed bulk flow. The simulation capabilities discussed in this work highlight the need to capture the production and stagnation processes inherent to plasma gun devices. Comprehensive simulations moving forward can build upon the framework introduced here and incorporate additional surface physics and kinetic effects to accurately quantify heating loading transferred to candidate materials through all phases of ELM transport and allow direct comparisons to degradation rates measured experimentally.

Although plasma guns operate over a range of conditions, the results reported in this paper show that flow properties and collisionality must be taken into account when extrapolating to fusion conditions. Specifically when exposing material targets in gun devices, the heat flux imposed on the surface must be measured at the target and not inferred from device or plume properties without considering collisional shielding. In addition, thermalization of particles not only changes the heating dynamics, it also reduces the characteristic velocity and thus the expected degradation rate of materials. Over time, modifications to the conventional QSPA configuration have been built and tested to better represent fusion conditions, such as the MK-200UG and MK-200CUSP. For fully consistent ELM simulation however, plasma guns will need to approach the flow conditions expected in disruptions themselves.

\ack
% Acnowledgements go here

This work is supported by the U.S. Department of Energy Grant No. DE-NA0002011. The authors also want to thank Hadland Imaging for graciously letting them borrow the camera utilized in this work and both Dr. Keith Loebner and Dr. Vic Miller for their assistance with setting up the Schlieren imaging experiment. Thomas Underwood also gratefully acknowledges the financial support of the National Defense Science and Engineering Graduate Fellowship Program.
\newpage
 
\bibliographystyle{iopart-num}
\bibliography{shockShieldCitations}

\end{document}